\documentclass[aps,pre,twocolumn,showpacs]{revtex4-1}
\usepackage{graphicx}
\usepackage{dcolumn,xcolor,ulem}
\usepackage{amsmath} 
\usepackage{amssymb}
\usepackage{amsfonts}
\usepackage{bm}
\usepackage[latin1]{inputenc}
\usepackage{fancyhdr}
\usepackage{amsmath} 
\usepackage{amssymb,txfonts}

\begin{document}

%Title of paper
\title{Syneresis and delayed detachment in agar plates}

\author{Thibaut Divoux}
\email[]{divoux@crpp-bordeaux.cnrs.fr}
\author{Bosi Mao}
\author{Patrick Snabre}
\email[]{snabre@crpp-bordeaux.cnrs.fr}
\affiliation{Université Bordeaux, Centre de Recherche Paul Pascal, UPR~8641, 115 av. Dr. Schweitzer, 33600 Pessac, France}

\date{\today}

\begin{abstract}
Biogels made of crosslinked polymers such as proteins or polysaccharides behave as porous soft solids and store large amount of solvent. These gels undergo spontaneous aging, called {\it syneresis} that consists in the shrinkage of the gel matrix and the progressive expulsion of the solvent. As a result, a biogel originally casted in a container often lose contact with the container sidewalls, and the detachment time is a priori difficult to anticipate since it may occur over variable time spans (from hours to days). Here we report on the syneresis phenomena in agar plates that consist in Petri dishes filled with a gel mainly composed of agar. Direct observations and speckle pattern correlation analysis allow us to rationalize the delayed detachment of the gel from the sidewall of the Petri dish. The detachment time $t^*$ is surprisingly not controlled by the mass loss as one would intuitively expect. Instead, $t^*$ is strongly correlated to the gel minimum thickness $e_{min}$ measured along the sidewall of the plate, and increases as a robust function of $e_{min}$ independently of the prior mass-loss history. Time-resolved correlation spectroscopy atypically applied to such weakly diffusive media gives access to the local thinning rate of the gel. This technique also allows us to detect the gel micro-displacements that are triggered by the water evaporation prior to the detachment, and even to anticipate the latter from a few hours. Our work provides observables to predict the detachment time of agar gels in dishes, and highlights the relevance of speckle pattern correlation analysis for the quantitative investigation of the syneresis dynamics in biopolymer gels.
\end{abstract}

\pacs{83.80.Kn,82.35.Pq,81.40.Cd,87.64.Cc}

\maketitle

\section{Introduction}
Biogels formed through the self-assembling of polymers such as polysaccharides or proteins are widespread in manufactured goods and biomimetic products \cite{Whistler:1973,Calvert:2009}. Fields of applications range from food engineering where biopolymers are used as gelling agents \cite{Mezzenga:2005}, to biotechnology where these gels commonly serve as growth media for microorganisms or as porous scaffold in tissue engineering \cite{Lee:2001,Rinaudo:2008}. Biogels exhibit a porous microstructure made of an interconnected network that is a priori efficient to retain solvents. However, these structures are often metastable. Indeed, the constituents experience attractive interactions and biogels spontaneously rearrange and shrink on durations ranging from hours to days depending on the ambiant relative humidity, leading to the progressive release of the solvent initially trapped. This phenomenon coined {\it syneresis} has been reported in biogels such as gelatin \cite{Kunitz:1928}, polysaccharide gels \cite{Matsuhashi:1990}, globular protein gels \cite{vanDijk:1984,Lodaite:2001,Mellema:2002}, organogels \cite{Wu:2009} and hydrogels from pNIPAM microgels \cite{Gan:2010}, and more generally in colloidal gels that display weak attractive interactions \cite{Teece:2010,Bartlett:2014}. If the latter category of gels has been the topic of numerous studies, only a handful of paper have reported quantitative measurements on the shrinkage dynamics of biogels, while the parameters controlling the detachment of a biogel from the preparation container stands as an open issue despite its oustanding practical importance.

Here we report on the syneresis process in commercial agar plates used as growth media for microorganisms or cells in routine diagnostic tests. These plates are usually incubated at constant temperature of about 35 to 40$^{\circ}$C for several hours. Excessive syneresis leads to the gel detachment from the sidewall of the Petri dish which makes it hard to assess any bacterial growth and invalidates the test. This simple detachment issue delays each year the analysis of thousands of diagnostics and costs a fair share of money to medical companies due to customer return.
The gelling constituant of these plates is agarose, a hydrophilic colloid extracted from seaweeds \cite{Matsuhashi:1990}, which formation and structural properties have been thoroughly investigated over the past 40 years \cite{Nijenhuis:1997,Lahaye:2001,Stanley:2006}. Insoluble in cold water, agar becomes soluble in boiling water and, once cooled down below $40^{\circ}$C, forms a thermoreversible gel that does not melt below $80^{\circ}$C. For concentrations above 1~\% (w/w) as it is the case for agar plates, the gel is formed through a competition between a spinodal demixing process and the association of molecules in double helices \cite{Feke:1974,Biagio:1996,Manno:1999,Emanuele:2004}. This process leads to a fibrous fractal-like microstructure that is controlled by the agarose concentration \cite{Normand:2000} and the thermal history \cite{Aymard:2001,Bulone:2004}. Such scale free microstructure reflects in their linear mechanical properties as the gel elastic modulus increases as weak power-law of the frequency \cite{Mohammed:1998,Barrangou:2006}. Agar gels behave as soft solids and display a brittle-like rupture scenario under large strains that involves macroscopic fractures, while considerable creep occurs under external stress before a delayed rupture \cite{Bonn:1998}.\\ 
Agar gels are also subject to ageing together with the spontaneous release of water. Such syneresis phenomenon is attributed to the contraction of the polymer network by a slow further aggregation of the helices \cite{Arnott:1974,Dea:1987} and is enhanced under external stress \cite{Matsuhashi:1990}, or low humidity conditions. Quantitative measurements of syneresis in agar gels merely consist in weighing the solvent-loss \cite{Boral:2010} and most of the current knowledge is limited to empirical laws regarding the influence of the gel composition: syneresis in agar gels goes roughly as the inverse square root of the polymer concentration \cite{Matsuhashi:1990}, while tunning the internal hydrophobicity of the gel by incorporating water-binding components such as sucrose \cite{Maurer:2012}, ester sulfate \cite{Stanley:2006}, xanthan \cite{Nordqvist:2011}, or locust bean gum \cite{Deuel:1950} may delay and/or prevent part of the water release. 
%To our knowledge, the influence of the competition between the water release and the gel adhesion to the wall upon the gel detachment scenario from its preparation container has received very little if no attention.   

%%%%%%%%%%%%
\begin{figure}[!t]
\centering
	\includegraphics[width=0.8\linewidth]{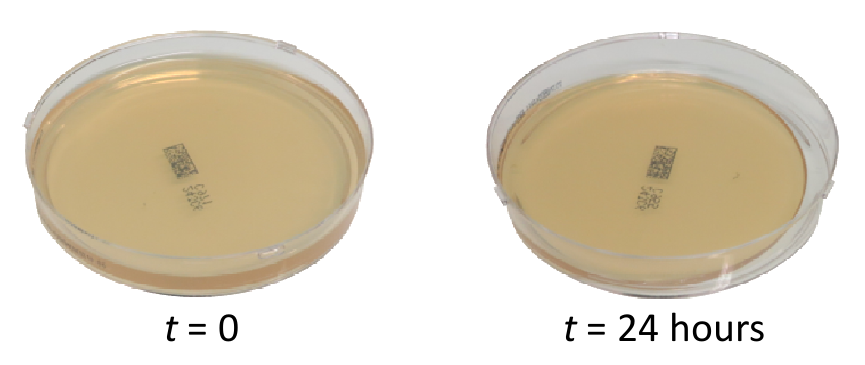}
\caption{Commercial agar plate from a fresh batch (left) and after a 24h incubation at 25$^{\circ}$C (right). The gel has detached from the sidewall on the right side of the dish. The scale is set by the dish diameter of 8.5~cm. 
\label{fig.agar}}
\end{figure} 
%%%%%%%%%%%%
	
In this article, we focus on the delayed detachment of agar gels in plastic Petri dishes that is triggered by the syneresis. The detachment is named {\it delayed} as it may occur from a few hours up to 30 hours, from the moment the plates are opened and incubated at constant temperature. Direct visualization and speckle pattern correlation experiments allow us to analyze the gel dynamics before its detachment from the sidewall of the dish, and to identify the key parameters controlling the detachment time $t^*$. Surprisingly, $t^*$ is not controlled by the water loss, as two gels of identical mass with different mass loss history may detach over very different timescales. Instead, the detachment time is strongly correlated to the gel thickness asymmetry along the plate periphery, and increases as a robust power law of the gel minimum thickness $e_{min}$ independently of the gel prior history. Such simple quantities constitute promising macroscopic observables to estimate and optimize the shelf life of commercial plates. Time-resolved correlation experiments further allow us to measure the gel local thinning rate and demonstrate that despite the gel detachment occurs in a sudden single step, it can be anticipated from a few hours by monitoring the gel micro-displacements while it is still in contact with the sidewall of the dish. 

\section{Materials and methods}

\subsection{Agar plate samples}
The samples consist in gamma-irratiated sterile agar plates (Fig.~\ref{fig.agar}) commercialized by BioM\'erieux as microbiological growth media\footnote[3]{Note that gamma irradiation is performed after gelation to sterilize the plates. To our knowledge, there are no studies dealing with the effect of post-gelation irradiation on the mechanical properties of agar gels. Nontheless, note that irradiation of agar prior to gelation is known to impact the gel mechanical properties by lowering the gel failure stress compared to non-irradiated samples \cite{Pietranera:2001}.}. Plates are made of a cylindrical plexiglas box (diameter 8.5~cm, height 1~cm) covered with a removable lid and are partially filled with an agar gel (1.5~\% w.t.) containing nutrients for bacterial growth that include peptones, sodium salts, etc.

%%%%%%%%%%%%
\begin{figure}[!t]
\centering
\includegraphics[width=0.95\linewidth]{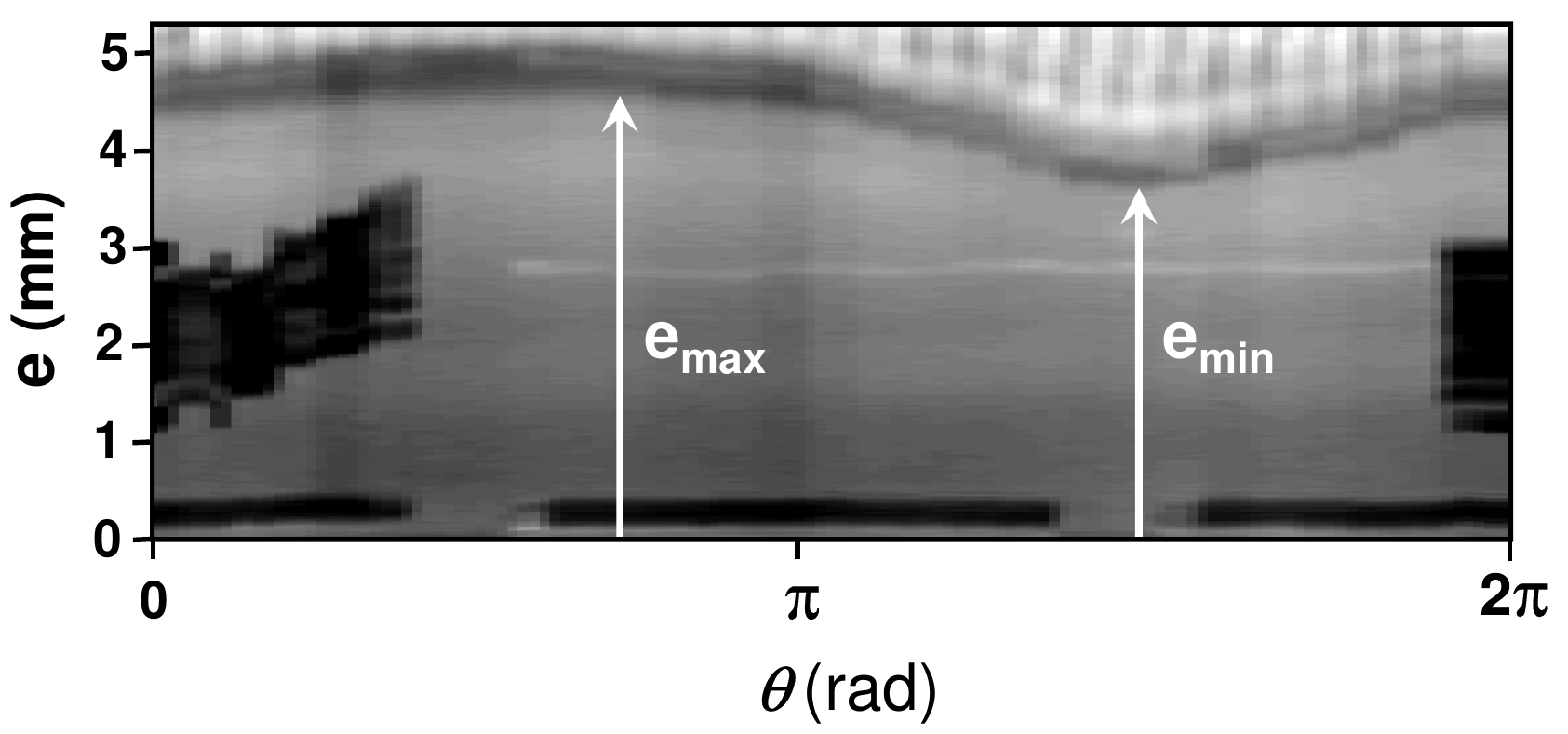}
\caption{Panoramic sideview of a commercial agar plate ($m=26$~g). The image is built by rotating the plate in front of a webcam and stacking consecutive vertical snapshots recorded at regular angle intervals. The representation $e(\theta)$ illustrates the variations of the gel thickness $e$ along the periphery of the Petri dish ($e_{min}=3.58\pm 0.03$~mm and $\delta e = e_{max} - e_{min}=1.02\pm 0.03$~mm). The image height is about 800~pixels and the black inscription is the plate serial number. 
\label{fig.thickness}}
\end{figure} 
%%%%%%%%%%%%
As a first key observation, one can notice that the gel thickness is not homogeneous, especially along the sidewall of the Petri dish. This point is of primary importance for the dynamics of the gel detachment from the dish as will be discussed in section \ref{macroscopic}. Prior to any use, each plate is weighted to determine the mass $m$ of gel it contains, and the gel thickness $e(\theta)$ along the sidewall of the Petri dish is measured by means of a webcam (Logitech HD c920) with an accuracy of $\pm$30~$\mu$m. In particular, we record the gel minimum and maximum thicknesses, resp. e$_{min}$ and e$_{max}$ (Fig.~\ref{fig.thickness}). 
For the samples investigated here, the gel mass $m$ ranges from 21 to 27~g, while the gel thickness asymmetry $\delta e \equiv e_{max}-e_{min}$ lies between 0.3 and 1.4~mm for a typical average thickness of about 4~mm. Both the gel weight $m$ and thickness $e(\theta)$ depend upon the gel casting process on the production line, and therefore are not control parameters in this study. This is why we have performed experiments on a large number of  plates to sample different values of $m$ and $e$. As a result of water exudation and evaporation, $m$ and $e$ progressively decrease up to the point the gel detaches from the lateral wall of the dish and further retracts [Fig.~\ref{fig.agar}~(right)] which marks the end of the product shelf life. Commercial agar plates comes in batches of 10 plates wrapped together. To test the reproducibility of our results, each experiment is repeated on several independent batches which exact number is given in the text.

\subsection{Direct visualization}

For each batch, the 10 plates are placed at the same level on top of a grid inside a programmable testing chamber (Binder MK53) that maintains a constant temperature of $(25.0 \pm 0.1) ^{\circ}$C and a relative humidity of about $50\%$. Plates are monitored over several hours (up to $\sim 40$ hours) with a webcam (Logitech HD c920) placed inside the chamber and used in timelapse mode with a frame rate of 1 image per minute.
%%%%%%%%%%%%
\begin{figure}[!t]
\centering
\includegraphics[angle=-90,width=1.\linewidth]{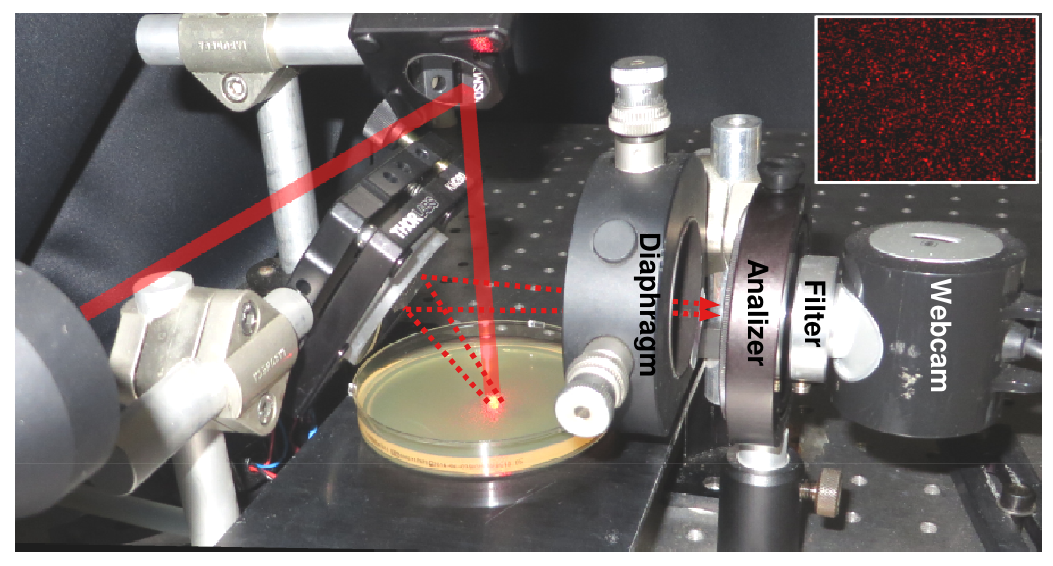}
\caption{Photography of the optical setup used to record the speckle pattern from the agar plate in the backscattering geometry. The laser beam trajectory has been sketched in red for the sake of clarity. Inset: typical speckle pattern.  
\label{fig.dws}}
\end{figure} 
%%%%%%%%%%%%

\subsection{Time correlation spectroscopy}
\label{setup}

The gel dynamics of a single plate can also be monitored through speckle pattern correlation in a room thermostated at (25.0$\pm$ 0.1)$^{\circ}$C and with a humidity level of about $50\%$. The optical setup pictured in figure~\ref{fig.dws}, consists in a linearly polarized laser beam (He-Ne Melles Griot gas laser, 20~mW at $\lambda=$632.8~nm) that is expanded to 5~mm with a collimator and directed perpendicularly to the sample by means of a mirror. The light is backscattered from the center region of the plate (illuminated volume $\sim$ 100~mm$^3$) and forms a speckle pattern on a low sensitivity CCD array (Webcam Phillips SPC900NC, 640$\times$480~pixels) an example of which is represented in the inset of figure~\ref{fig.dws}. To remove the ambient light, an interference filter is placed in front of the detector and for each experiment, the angular position of the plate and the diaphragm aperture are tuned to spread the intensity range over the whole accessible grayscale of the detector, and obtain an autocorrelation radius ($\sim$ speckle grain radius) of about 3 pixels. We have checked that the results reported here do not depend on the exact position of the enlightened volume in the sample.   

As water evaporates, the gel thins and the speckle pattern changes. The degree of correlation between two speckle images separated by a lag time $\tau$ is determined by the ensemble-average intensity correlation function $g_2(t,\tau)$ defined as follows:
\begin{equation} 
g_2(t,\tau) = \frac{\langle I_p(t)\cdot I_p(t+\tau )\rangle_p}{\langle I_p(t)\rangle_p \cdot \langle I_p(t+\tau) \rangle_p}
\label{eq.1}
\end{equation}
where $I_p$ denotes the brightness level of pixel $p$ and $\langle ... \rangle_p$ an ensemble average over all the pixels \cite{Cipelletti:2003}. The correlation function is further normalized into a function noted $g^*_2(t,\tau)$ that verifies the condition $g^*_2(t,\tau=0)=1$. Here, $g^*_2(t,\tau)$ is computed at a frame rate of 10~Hz over a lag time $\tau$ ranging from $0$ to $1$~min. 
%while for $\tau >1$~min, $g^*_2(t,\tau)$ is only computed every minute. 
Speckle patterns are processed in real time by means of a custom-made java plug-in with the NIH Image processing package, and speckle images are only saved every minute for a lag time $\tau=0$ to avoid storing too large amount of data \cite{Snabre:2009}. 

\begin{figure}[!t]
\centering
\includegraphics[width=1.\linewidth]{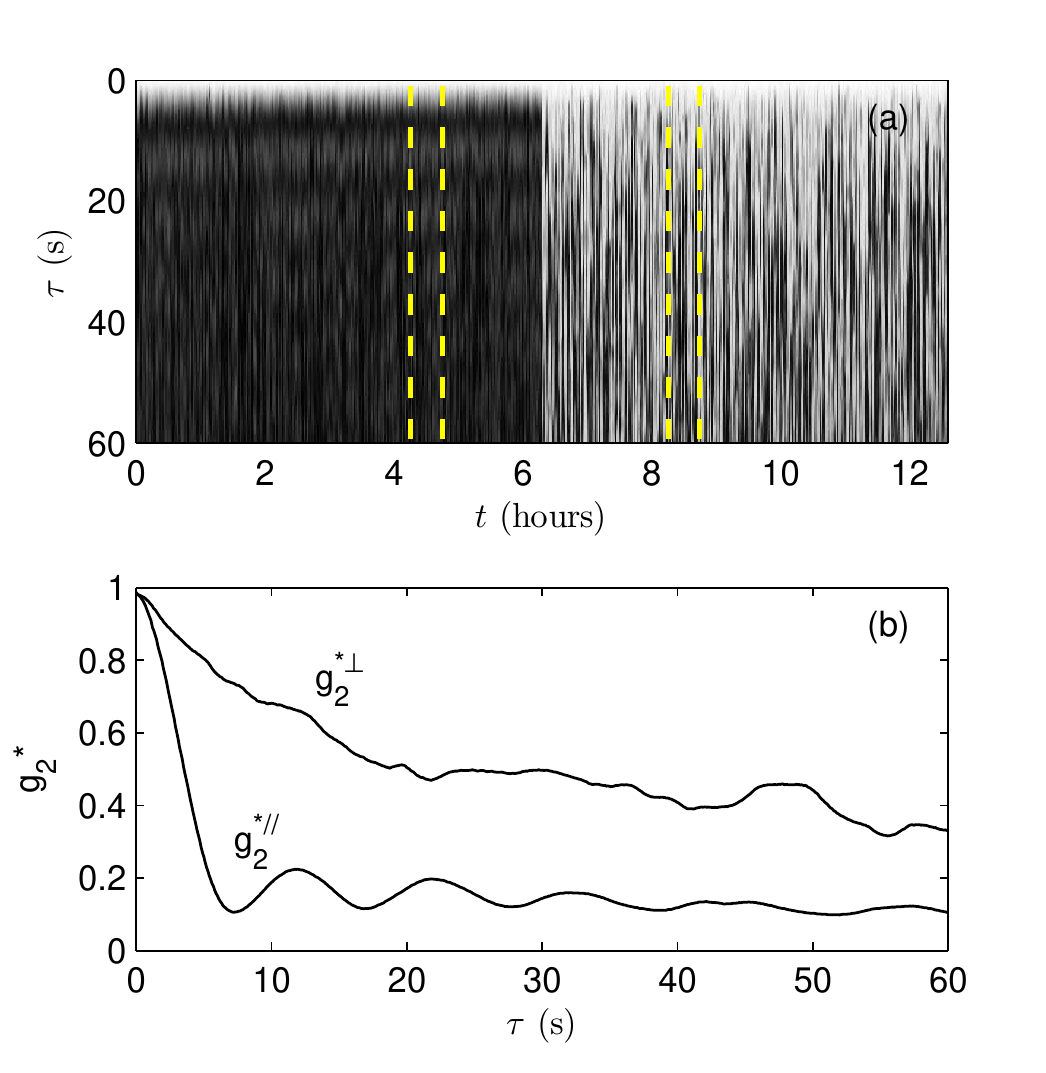}
\caption{(Color online) (a) Lag-time temporal diagram of the intensity correlation function $g^*_2(t,\tau)$ coded in grayscale as a function of the lag time $\tau$, and the experimental time $t$. The first half of the experiment is performed with parallel polarizers ($t<6.3$~hours) and the second half of the experiment with crossed polarizers ($t\geq 6.3$~hours). The detachment of the gel from the sidewall of the dish occurs later at $t=16$~hours (data not shown). (b) Intensity correlation function $g^*_2$ vs. the lag time $\tau$ extracted from (a) at $t=4.5$~hours (parallel configuration) and at $t=8.5$~hours (crossed configuration) and averaged over a time window of $\Delta t=30$~min [duration enclosed between the yellow doted lines in (a)]. The gel  characteristics are the following: $m=26$~g, $e_{min}$=3.74~mm and $\delta e$=0.75~mm.  
\label{fig.speckleMethod}}
\end{figure}

The gel is a weakly scattering media and the speckle pattern is mainly produced by the interference of the light beam initially polarized, that is either reflected by the air/gel interface and/or by the gel/dish bottom interface. The polarization of the light reflected by the air/gel interface is unchanged, whereas the polarization of the light reflected by the gel/dish bottom interface is modified, as the Petri dish is made out of a birefringent material. To take advantage of this situation, an analyzer is placed in front of the webcam (Fig.~\ref{fig.dws}) and is oriented either parallel or perpendicular to the original polarisation direction. To illustrate the speckle evolution in each of these two configurations, we report an experiment on an agar plate which first (second resp.) half is performed in the parallel (perpendicular resp.) configuration [Fig.~\ref{fig.speckleMethod}(a)].

In the parallel configuration, the speckle results from the interference of the light reflected at the air/gel interface and at the gel/bottom plate interface and its decorrelation is mainly due to the gel thinning. Quantitatively, the correlation function $g_2^{*\varparallel}$ reported in figure~\ref{fig.speckleMethod}(b) decreases rapidly over a timescale $\tau_d^{\varparallel} \simeq 5$~s and further displays a periodic modulation $\tau_M^{\varparallel}$ of about 8~s that is also clearly visible in the dynamics of the lag-time temporal diagram of the correlation function [Fig.~\ref{fig.speckleMethod}(a), for $t<6.3$~h]. 
On the one hand, the short time decorrelation is induced by any change of $\lambda/2$ in the two optical paths. Indeed, a constant evaporation rate $\dot m =0.7$~g/hour at 25$^{\circ}$C (see table~1 in section~\ref{macroscopic}) of a cylindrical gel of radius $R=4.2$~cm and density $\rho=1000$~kg/m$^3$ leads to $\tau_d^{\varparallel}=\lambda \rho \pi R^2/(4\dot m) \simeq 4.5$~s in agreement with the initial decay of $g_2^{*\varparallel}$ in figure~\ref{fig.speckleMethod}(b). 
On the other hand, the modulation of the correlation function is caused by the constructive interference due to the gel thinning. 
%of the light that is reflected at the air/gel interface and the light that goes through the gel and is reflected at the bottom of the plate. 
Two successive maximum of the intensity correlation function $g_2^{*\varparallel}$ correspond to a change $\Delta$ in the optical paths of photons either reflected at the air/gel interface or at the dish bottom such as $\Delta=2nd=\lambda$, where $d$ denotes the decrease in the gel thickness, and $n$ stands for the refractive index of the gel very close to that of water ($n\simeq1.33$). As the air/gel interface goes down at a velocity $v=\dot m/(\rho \pi R^2)=d/\tau_M^{\varparallel}$, we can estimate the period of the oscillations to be $\tau_M^{\varparallel} = \lambda/(2nv) \simeq 7$~s, in agreement with the modulation period of $g_2^{*\varparallel}$ [Fig.~\ref{fig.speckleMethod}(b)].

%%%%%%%%%%%%
%%%%%%%%%%%%

In the perpendicular configuration, the webcam barely receives the light reflected at the air/gel interface which polarization is orthogonal to its initial value, and is now filtered by the analyzer. Therefore, in this configuration the speckle pattern is far less sensitive to the displacement of the air/gel interface and it decorrelates much more slowly than in the parallel configuration [Fig.~\ref{fig.speckleMethod}(b)]. Nonetheless, the decorrelation still occurs and results from 
%($i$) the gel thermal motion, 
($i$) the micro-displacement of the gel inside the dish\footnote[4]{Note that the displacement responsible for the speckle decorrelation is due to the micro-displacement of the gel and also to the deformation of the Petri dish under the forces exerted by the gel that contracts, because of the water evaporation. We have checked that the use of a rigid container (e.g. a Petri dish made of glass instead of plastic) somewhat dampens the temporal fluctuations of the speckle pattern observed in perpendicular configuration [see Fig.~1 in the supplemental material].} while the gel thins but remains macroscopically in contact with the sidewall, and from ($ii$) the local changes in the surface topography or in the orientation of the air/gel interface induced by the aforementioned displacements.

%%%%%%%%%%%%
\begin{figure}[!t]
\centering
\includegraphics[width=\linewidth]{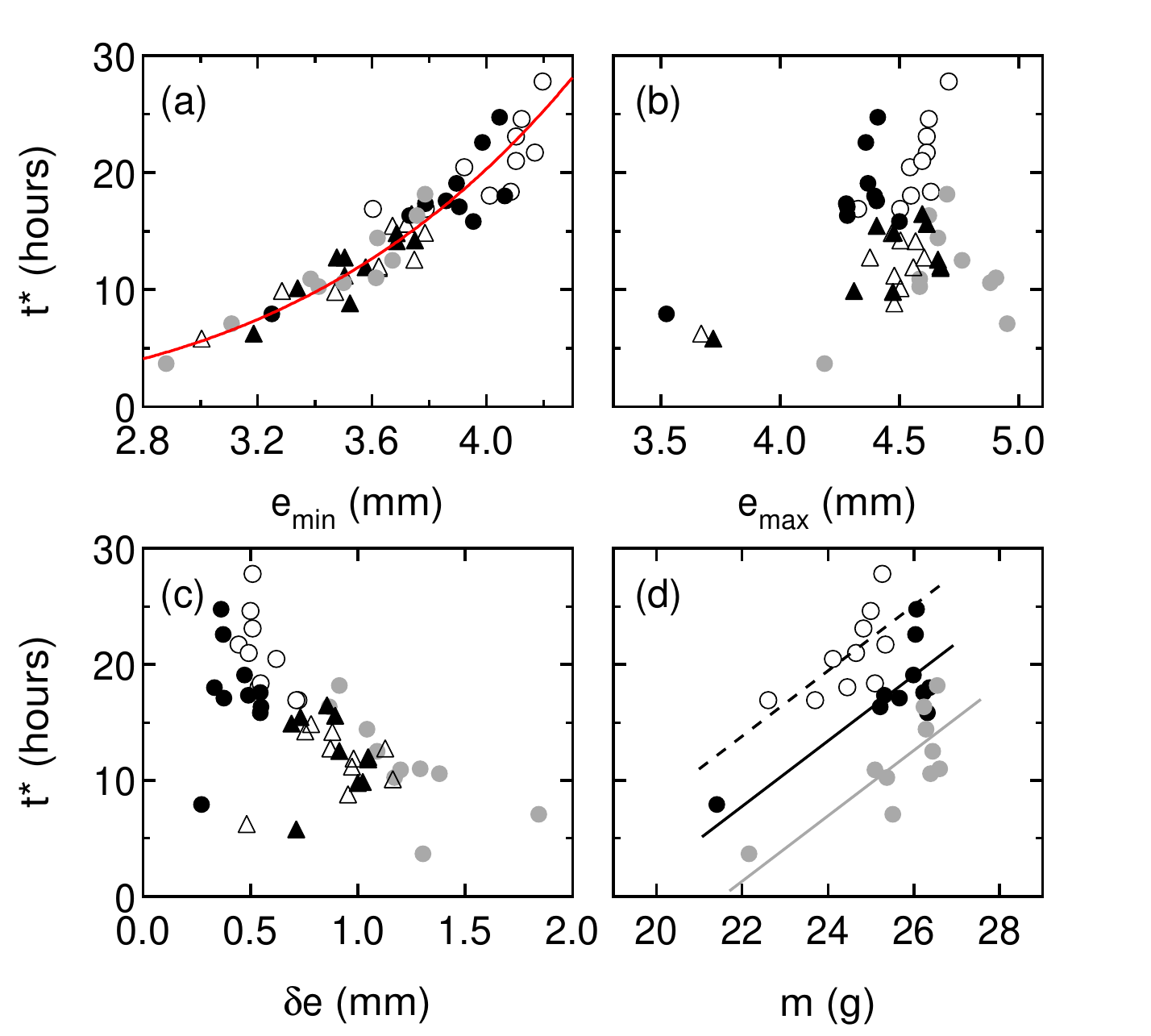}
\caption{(Color online) Detachment time $t^*$ of the gel from the sidewall of the Petri dish plotted vs the value measured at $t=0$ of (a) the gel minimum thickness $e_{min}$, (b) the gel maximum thickness $e_{max}$, (c) the thickness asymmetry $\delta e=e_{max}-e_{min}$ and (d) the gel weight $m$. The red curve in (a) is the best power-law fit of the data that passes through the origin: $t^*=4\cdot 10^{-2} e_{min}^{4.5}$. The lines in (d) are guides for the eye. Symbols ($\circ$), (\textcolor{white!35!black}{$\bullet$}) and (\textcolor{black}{$\bullet$}) stand for two independent batches of 10 plates each used in direct visualization experiments, while ($\triangle$) and ($\blacktriangle$) stand for three independent batches of 10 plates for which $t^*$ was determined by correlation spectroscopy experiments (see section~\ref{TRC}).}
\label{fig.3}
\end{figure} 
%%%%%%%%%%%%

The micro-displacements of the gel within the dish are triggered by the successive relaxations of the internal stress. The latter progressively builds up as the gel tends to contract due to water evaporation, but sticks to the sidewall. The average internal stress increases up to the moment the contraction forces overcome the adhesion forces between the gel and the sidewall leading to the sudden detachment. Therefore, in the perpendicular configuration, the temporal evolution of the speckle reflects the gel dynamics well before the detachment and allows us to infer the micro-displacement of the gel during the syneresis. 
Furthermore, in this configuration one can see in figure~\ref{fig.speckleMethod}(a) (for $t\geq 6.3$~h) that the speckle decorrelates over a timescale that varies considerably, from seconds to minutes, which suggests that the occurrence of these micro-displacements is intermittent. A more detailed analysis of these fluctuations reported in section~\ref{TRC} confirms this interpretation and reveals that TRC allows us to detect precursors to the gel detachment from sidewall of the dish.

\section{Results}

\subsection{Macroscopic approach}
\label{macroscopic}

In a first series of experiments, a batch of 10 plates without their lids is placed in a thermoregulated chamber (T=25$^{\circ}$C) and left at rest. Water evaporates from the agar plate and the gel stay still for several hours before suddenly detaching from the sidewall of the Petri dish at a time noted $t^*$ [See movie~1 in the supplemental material]. After detaching, the gel shrinks and although the gel may lose up to 40~\% of its initial weight during the syneresis, we observe no sign of failure or fracture at any time\footnote[5]{Note that, pushing the experiments forward, the gel turns into a thin and dry buckled pancake that can be rejuvenated by adding water. The gel then swells and recovers its initial cylindrical shape, occupying the whole Petri dish after a few hours (data not shown).}.

%%%%%%%%%%%%
\begin{figure*}[!t]
\centering
\includegraphics[width=0.85\linewidth]{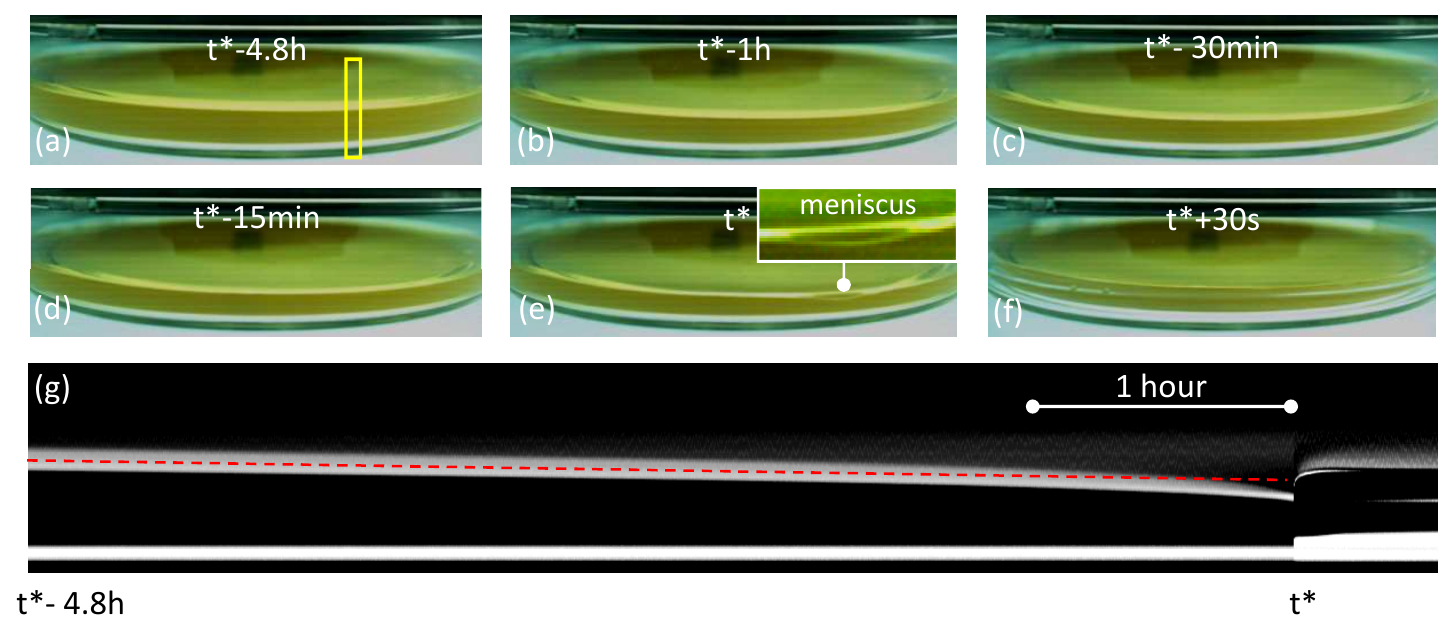}
\caption{(a)-(f) Sideview images of a gel in a Petri dish during the syneresis, taken at different times before and after the detachment of the gel from the sidewall of the dish. Note in (e) the appearance and the growth of a lens shaped meniscus at the exact location where the detachment will occur at $t^*=4.8$~hours. (f) Spatiotemporal diagram of the gel thickness $e(t)$ as a function of time and computed in the region of interest emphasized by a yellow frame in (a) which corresponds to the detachment location. As shown by the red dashed line (guide for the eye), the gel thickness decreases linearly with time in this region, up to one hour before the detachment. Then, the gel thinning speeds up concomitantly with the growth of a lens shaped meniscus visible in (e) before the gel detachment. The gel characteristics are: $m=23.8$~g, $e_{min}=3.15$~mm and $\delta e=1$~mm. 
\label{fig.macro}}
\end{figure*} 
%%%%%%%%%%%%

To identify the control parameters of the gel detachment during the early stage of the syneresis, we report in figure~\ref{fig.3} the evolution of the gel detachment time versus the gel initial characteristics. As a key result, $t^*$ is observed to increase as a power law of the gel minimum thickness $e_{min}$, measured along the sidewall of the Petri dish [Fig.~\ref{fig.3}(a), ($\circ$), (\textcolor{white!35!black}{$\bullet$}) and (\textcolor{black}{$\bullet$})]. This relation is robustly verified over 30 plates of different initial weights and mass loss history. By contrast, the detachment time is not correlated to the gel maximum thickness [Fig.~\ref{fig.3}(b)]. Moreover, the detachment time also increases with the mass of the gel [Fig.~\ref{fig.3}(d)]. Nonetheless, if the data obtained with the same batch can be described with a single increasing function of $m$, two gels with the same initial mass but taken from different batches may detach over very different timescales. This last observation is likely related to the fact that plates from different batches may have been exposed to different storage conditions and may have lost various amount of water since their production. Last but not least, the detachment time also appears to be correlated to the amplitude of the variations of the gel thickness quantified by $\delta e = e_{max}-e_{min}$, but to a lesser extent as the data are more scattered than for $t^*$ vs $e_{min}$ [Fig.~\ref{fig.3}(c)]. One should nonetheless keep in mind that gels with larger thickness asymmetry are more likley to detach sooner from the sidewall.  

Beyond the detachment time, we have monitored the evolution of the gel thickness and the weight of the Petri dishes stored in the thermoregulated chamber. The experiments show that the water loss increases linearly with time, while the gel detachment does not affect the mass-loss rate (see Fig.~2 in the supplemental material). This result is robust and experiments performed at larger temperatures simply lead to similar results with larger loss-rates (Table~\ref{table.1}). Moreover, the average thickness of the gel decreases linearly in time in agreement with the evolution of the mass loss, while the thickness asymmetry $\delta e$ remains about constant up to the detachment (see Fig.~3 in the supplemental material). The latter result urges to have a closer look at the gel dynamics where the gel thickness is minimal.

The temporal evolution of the gel thickness, at the very location where the detachment takes place is reported in figure~\ref{fig.macro}~(a)--(f). One can see that the gel also thins linearly with time at this specific location. However, about one hour before the detachment, the thinning speeds up in this spot as evidenced in the spatio-temporal diagram displayed in figure~\ref{fig.macro}(f) and the growth of a lens-shaped meniscus [Fig.~\ref{fig.macro}(e)] finally leads to the detachment of the gel from the sidewall. This observation appears as robust and confirms that this specific area of the gel in contact with the sidewall of the dish plays a key role in the detachment process and necessitate more sensitive measurements.

\begin{table}[!b]
\begin{center}
\begin{tabular}{|ccc|ccc|}
 \hline
 & Temperature ($^{\circ}$C) &  & & $\delta \dot m$ (g/hour) & \\
 \hline
 & 20 & & & $0.50 \pm 0.02$ & \\
 & 30 & & & $0.86 \pm 0.03$ & \\
 & 40 & & & $1.81 \pm 0.08$ & \\
 \hline
\end{tabular}
\end{center}
\caption{Mass loss rate $\delta \dot m$ determined at three different temperatures $T$. Each value is the result of a linear fit of $\delta m (t)=m(0)-m(t)$ for 6 different plates extracted from three different batches. Measurements are detailed in figure~2 in the supplemental material.}
\label{table.1}
\end{table}

\subsection{TRC study of the syneresis}
\label{TRC}

To get deeper insights on the gel dynamics and in particular on the local displacements that the gel experiences before the detachement, we performed a series of speckle pattern correlation experiments. An agar plate is placed at T=25$^{\circ}$C in the custom made optical setup described in section~\ref{setup} which is set in the perpendicular configuration to filter the light reflected at the air/gel interface and avoid the short time decorrelation of the speckle. The gel dynamics during the water evaporation is quantified by computing the intensity correlation function $g^{*\bot}_2(t,\tau)$ over a lag time $\tau=1$~min repeatedly, during 10 to 30 hours (including during and after the macroscopic detachment).

%%%%%%%%%%%%
\begin{figure*}[!t]
\centering
\includegraphics[width=0.85\linewidth]{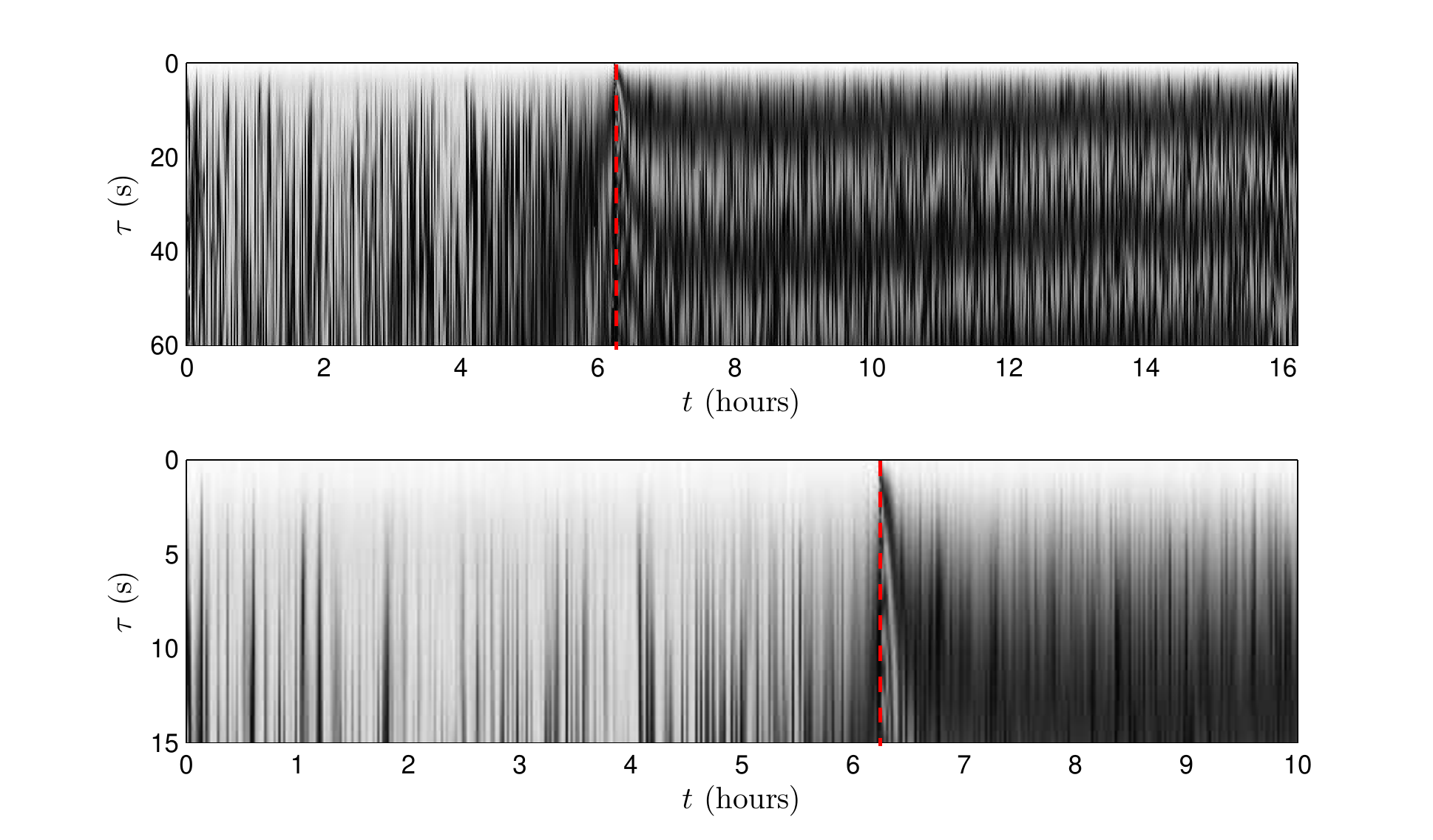}
\caption{(a) Lag-time temporal diagram of the intensity correlation function $g^{*\bot}_2(t,\tau)$ coded in grayscale and measured in the perpendicular configuration, as a function of the lag time $\tau$ and the experimental time $t$. The vertical red dashed line at $t \equiv t^*=6.25$~hours indicates the time at which the gel detaches from the sidewall of the Petri dish. For $t\leq t^*$, the speckle decorrelates over a fluctuating timescale due to the micro-displacements of the gel in contact with the sidewall of the dish. For $t>t^*$, the gel is free to contract and the speckle decorrelates more rapidly. The modulation of $g^{*\bot}_2$ of about $20$~s is related to the gel thinning (see text). (b) Zoom over the early stage of the lag-time temporal diagram, for short lag times. The gel characteristics are: $m=21.64$~g, $e_{min}=3.18$~mm and $\delta e=0.48$~mm. 
\label{fig.spatio}}
\end{figure*} 
%%%%%%%%%%%%

Figure~\ref{fig.spatio}(a) displays the lag-time temporal diagram of the correlation function $g^{*\bot}_2(t,\tau)$ coded in grayscale. One can see that $g^{*\bot}_2(t,\tau)$ exhibits two different regimes that point toward a dramatic event at a specific timescale $t=t^*$ [red dashed line in fig.~\ref{fig.spatio}]. To compare this timescale to the detachment time measured by direct visualization experiments, we have carefully repeated the experiment for twenty agar plates from two batches, and the time $t^*$ determined through TRC measurements is systematically reported vs. $e_{min}$ in figure~\ref{fig.3} [Symbols ($\triangle$) and ($\blacktriangle$)]. All the data points fall on the exact same power-law dependence as found in section~\ref{macroscopic}, which demonstrates that the time $t^*$ derived from the local measurements coincides with the macroscopic detachment time identified by direct visualization experiments. 

Before and after $t^*$, the correlation function displays different behaviors that reflect two distinctive dynamics of the gel. 
For $t<t^*$, the gel sticks to the sidewall of the dish, and due to water evaporation, the air/gel interface is streched and flat. In the perpendicular configuration, the light reflected at this interface is mostly filtered by the analyzer (section~\ref{setup}) and the correlation function reflects the micro-displacement of the gel that we discuss in more details in the following paragraph. 
For $t>t^*$, the gel has detached from the sidewall and is free to contract. The periodic modulation of the correlation function of about $\tau \simeq 25$~s [Fig.~\ref{fig.spatio}(a) for $t>t^*$] is due to the gel thinning\footnote[6]{Indeed, two successive maximum of the intensity correlation function $g^{*\bot}_2$ correspond to a change of $\Delta=\lambda/[2(n-1)]$ in the optical paths of the photons reflected at the gel/dish bottom interface. Estimating the gel thinning speed $v$ from the mass loss rate $\dot m$ (see section~\ref{setup}), one may derive the modulation period to be $\tau_M^{\bot }=\lambda/[2(n-1)v] \simeq 30$~s, which is consistent with the experiments.}.
This interpretation is further supported by a supplemental experiment performed on a fresh plate which gel has been carefully detached from the sidewall of the dish by means of a cutter blade, before the start of the experiment (see Fig.~4 in the supplemental material). In that case, the correlation function displays, from $t=0$, a modulation very similar to the one reported for a standard gel after it has detached (i.e. for $t>t^*$). 

%%%%%%%%%%%%%%%%%   
\begin{figure}[t]
\centering
\includegraphics[width=1.05\linewidth]{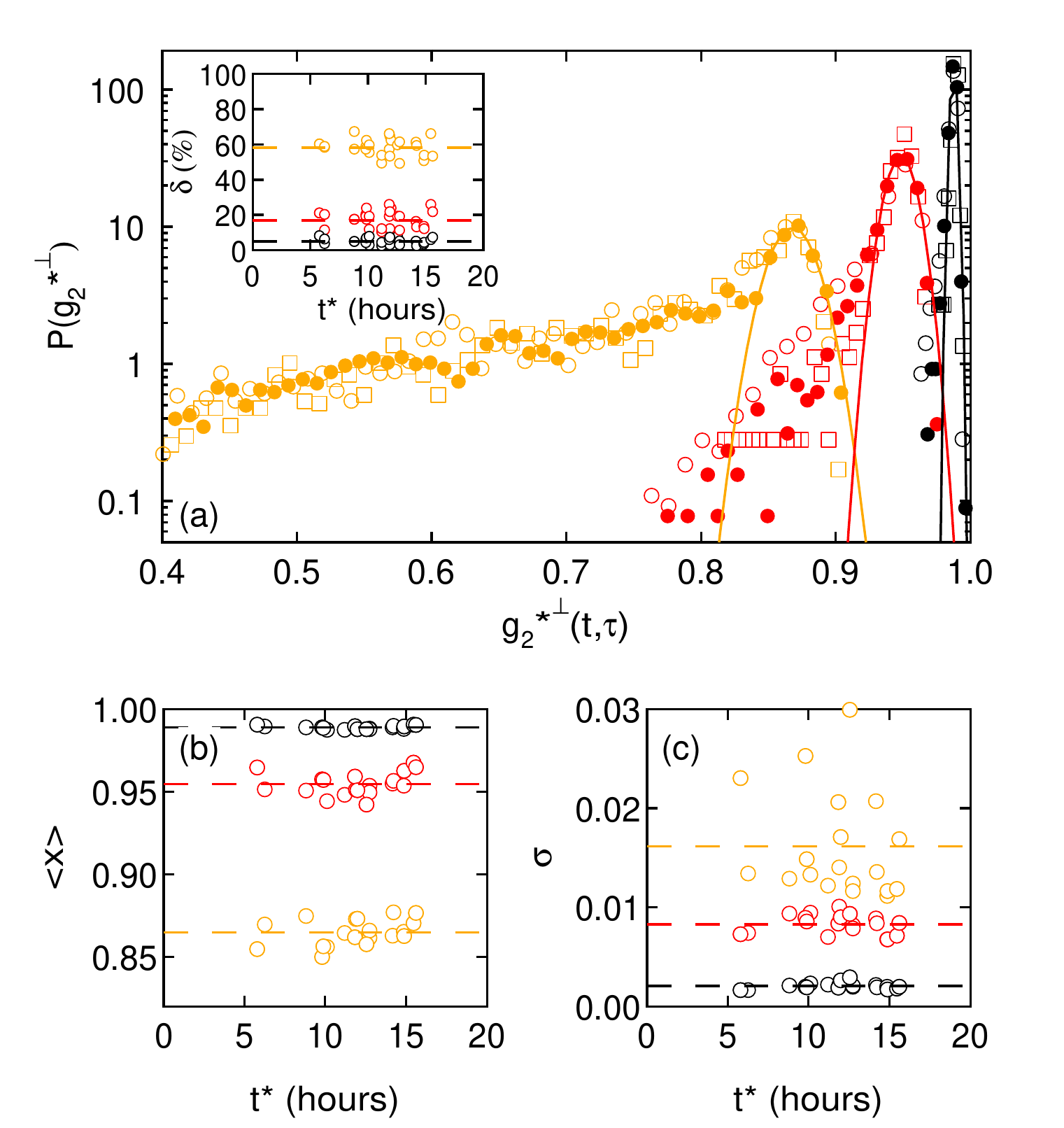}
\caption{(Color online) (a) Probability distribution function of $g^{*\bot}_2(t,\tau)$ computed at $\tau=0.5, 2$ and 10~s from right to left, for three different plates (same symbols as in fig.~\ref{fig.decor}). For ($\square$) the distribution is computed for $t$ ranging between $20$~min and 3~hours, while for ($\circ$) and $(\bullet$) the distribution is computed between $20$~min and up to $8$~hours. Inset: fraction $\delta$ of data values that lays outside the Gaussian fit of the data reported in (a) vs the detachment time. (b) Mean value $\langle x\rangle$ and (c) variance $\sigma$ extracted from Gaussian fits for twenty different plates and plotted vs. the detachment time. Horizontal dashed lines stand for the mean values of the data.   
\label{fig.PDF}}
\end{figure} 
%%%%%%%%%%%%

Let's turn now to the gel micro-displacements that take place before the detachment by having a closer look at $g^{*\bot}_2(t,\tau)$ for $t<t^*$ [Fig.~\ref{fig.spatio}(b)]. As mentioned in section~\ref{setup}, in the crossed polarizers configuration, the speckle decorrelation results from the micro-displacements of the gel within the dish that drive the speckle dynamics at short lag times ($\tau\leq 30$~s, see Fig.~5 in the supplemental material), and from the local subsequent changes in the topography or orientation of the air/gel interface that lead to the complete decorrelation of the speckle at longer times. Therefore, here we focus on the temporal evolution of $g^{*\bot}_2(t,\tau)$ for $\tau\leq 30$~s. We have computed for three different plates, the probability distribution function (PDF) of $g^{*\bot}_2(t,\tau)$ pictured in semi-logarithmic scale in figure~\ref{fig.PDF}(a) at $\tau=0.5, 2$ and 10~s. 
The PDF remains nearly Gaussian for short lag times $\tau\leq 0.2$~s, while $g^{*\bot}_2(t,\tau)$ explores smaller values for increasing values of the lag time, and the distribution $P(g^{*\bot}_2)$ develops an exponential tail [Fig.~\ref{fig.PDF}(a)]. The data located to the right of the maximum are fitted by a Gaussian function which center $\langle x\rangle$ and variance $\sigma$ are reported in figures~\ref{fig.PDF}(b) and (c) for twenty plates. Both parameters are independent of the detachment time of the gel which demonstrates that the gel dynamics on short time scale is thermally controlled by water evaporation and not related to the sudden macroscopic detachment. The latter is more likely the result of the accumulation of micro-displacements. This result also suggests that the Gaussian part of the distribution is related to random micro-displacements that do not lead to the gel relaxation on short timescales. In this framework, the growth of a non-Gaussian tail to the distribution of $g^{*\bot}_2$ for $\tau\geq 2$~s is related to irreversible micro-displacements of the gel relative to the dish that relax adhesion forces between the gel and the sidewall.
The amount of these micro-displacements for a given lag time $\tau$ can be assessed by computing the fraction of data points $\delta$ in the PDF $P(g^{*\bot}_2)$ that lays outside the Gaussian fit. One can see in the inset of figure~\ref{fig.PDF}(a) that $\delta$ neither depends on the plate nor on the gel detachment time, and increases with the lag time to reach $\sim$60\% at $\tau=10$~s. 
%This timescale is comparable to the modulation period $\tau_M^{\varparallel}$ measured with the parallel configuration in section~\ref{setup} that is associated to the gel thinning [Fig.~\ref{fig.speckleMethod}(b)], which confirms that these gel micro-displacements are driven by the water evaporation.
This timescale is comparable to the half modulation period $\tau_M^{\perp}/2\simeq 12$~s measured during gel thinning in the perpendicular configuration, and after the detachment from the dish has occured [Fig.~\ref{fig.speckleMethod}(a), $t>t^*$] which confirms that these gel irreversible micro-displacements are driven by water evaporation.

\begin{figure}[t]
\centering
\includegraphics[width=1.05\linewidth]{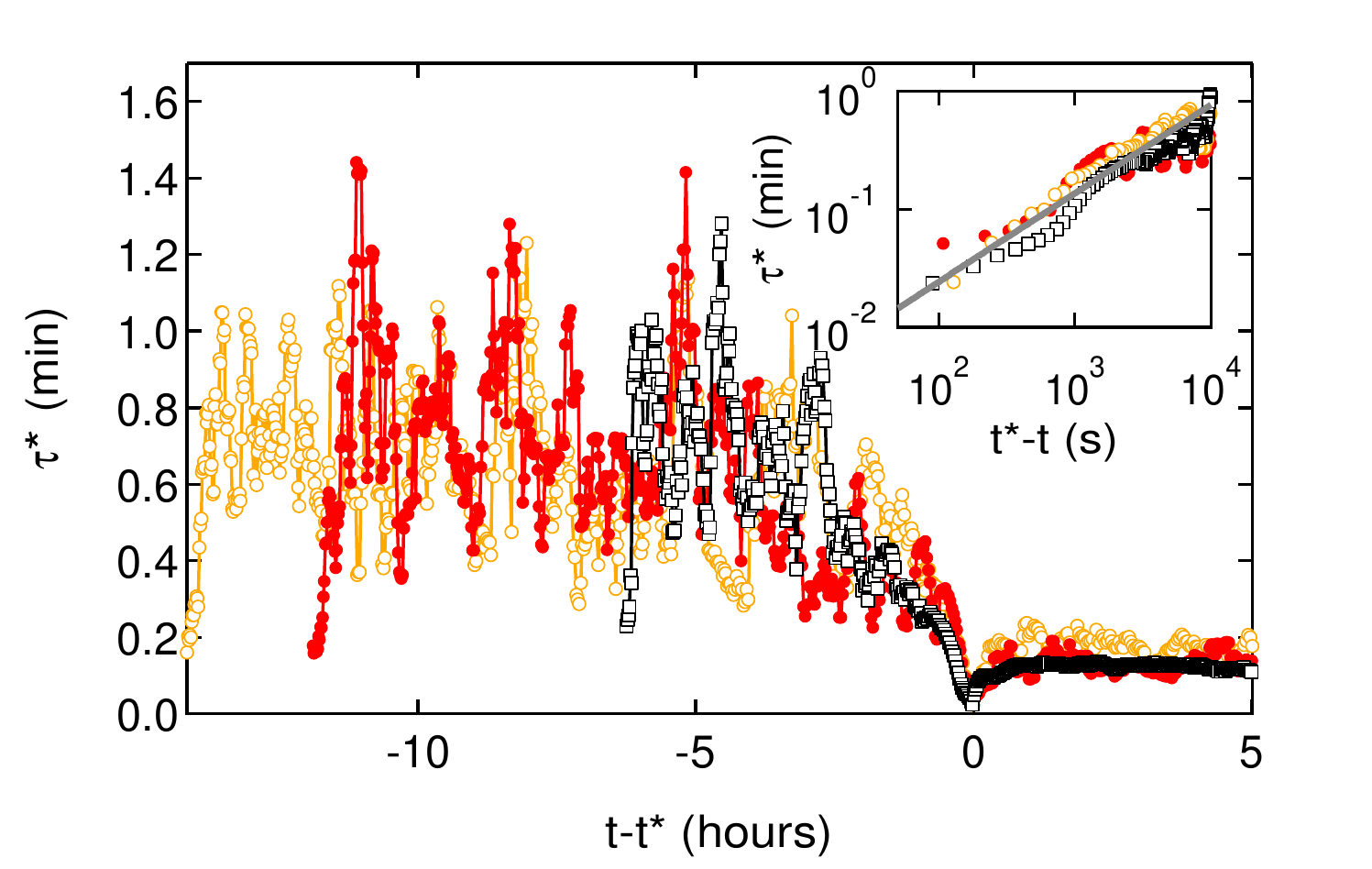}
\caption{(Color online) Evolution of the decorrelation time $\tau^*$ defined as $g^{*\bot}_2(t,\tau^*)=g^{*\bot}_2(t,0)/e$ vs. $t-t^*$, the elapsed time centered on the detachment time $t^*$. Each point is an average over $\Delta t=10$~min. Symbols stand for different plates: ($\square$) is the same plate as reported in fig.~\ref{fig.spatio} which gel detaches at $t^*=6.248$~hours, while (\textcolor{yellow}{$\circ$}) and (\textcolor{red}{$\bullet$}) display detachment times respectively of $t^*=14.16$~hours and 11.89~hours. Inset: Same data replotted vs. $t^*-t$ in logarithmic scale in the vicinity of the detachment. The line is the best power-law fit of the data: $\tau^* \propto (t^*-t)^{0.75}$.  
\label{fig.decor}}
\end{figure}

Finally, the last but not least remarkable feature of the lag-time temporal diagram pictured in figure~\ref{fig.spatio}(a), is that the decorrelation rate of the speckle pattern strongly increases about two hours before the gel detachment. The decorrelation time defined as the lag time $\tau^*$  for which $g^{*\bot}_2(t,\tau^*)=g^{*\bot}_2(t,0)/e$ is plotted in figure~\ref{fig.decor} as a function of the elapsed time $t-t^*$ for three plates that display very different detachment times $t^*$ ranging from 6 to 14~hours. For $t^*-t \gtrsim 2.5$~h, $\tau^*$ exhibits large fluctuations around a constant mean value of about 0.7~min in agreement with the fact that the evaporation rate is constant and the gel thickness decreases linearly in time (section~\ref{macroscopic}). For $t^*-t \lesssim 2.5$~h, $\tau^*$ decreases as a robust power-law that scales as $(t^*-t)^{0.75}$ independently of the detachment time (inset of Fig.~\ref{fig.decor}). This result shows that the number and/or the amplitude of the gel micro-displacements within the dish increases in a robust fashion up to the detachement. Such increase in the gel activity is also probably related to the formation and the growth of the lens-shaped meniscus in the vicinity of the sidewall at the location where the detachment later takes place [Fig.~\ref{fig.macro}(e) and (g)].
Furthermore, such power-law scaling is reminiscent of the finite time singularity reported for the breakup of liquid droplets \cite{Brenner:1997} and could be the signature of the fission at $t=t^*$ of the liquid film located between the gel and the sidewall. Indeed, we checked by means of a homemade compression cell that small deformations of about 3 to 5\% are sufficient to trigger the release of water from the gel. Therefore, one can easily imagine that the stresses generated by evaporation between the gel and the sidewall may form such liquid film. This last point is left for future work.

%%%%%%%%%%%%
%%%%%%%%%%%%

\section{Discussion and outlooks}

%\subsection{Discussion}

Agar plates display syneresis and water evaporation which lead to the delayed detachment of the gel from the sidewall of the dish. Water loss appears as a continuous process that is not affected by the gel detachment, and which rate increases for increasing temperatures. The gel thickness decreases linearly up to the detachment which occurs at the exact angular location where the gel thickness was originaly minimal. As a key result confirmed by both local and global measurements, the detachment time itself is strongly correlated to the gel minimum thickness, independently of the prior mass-loss history of the gel. It suggests that the detachment is governed rather by the adhesion of the gel to the dish, than by the water loss. Speckle pattern correlation experiments give access to the local thinning rate of the gel that is about a few $10$~nm/s, and show that the gel experiences micro-displacements induced by the water loss. As water evaporates, the gel tends to contract but remains in contact with the sidewall of the dish due to weak adhesion forces. The stress builds up which triggers irreversible micro-displacements up to the sudden detachment at $t=t^*$. The amplitude and/or the number of these displacements increases dramatically a few hours before $t^*$ which diminishes the decorrelation time of the speckle and makes it possible to anticipate the detachment by TRC analysis.

However, the exact location and the spatial extent of the gel micro-displacements inside the dish remains an open question: they could be located at the bottom of the dish or at the sidewall. Furthermore, if there is no doubt that the weak adhesion forces between the gel and the sidewall of the dish strongly impact the detachment time, the exact role of the friction properties between the gel and the dish bottom stands as an open issue. Future experiments will involve simultaneous TRC experiments in various region of interest and Photon Correlation Imaging\cite{Secchi:2013} to map the gel displacements over the dish bottom prior to the detachment. 

Finally, a local scenario of the syneresis at the scale of a single pore of the gel and in presence of evaporation is still lacking\cite{Scherer:1989}. It is likely that the water is released homogeneously at the nanoscale, and does not contribute to the intermittent dynamics of the speckle observed with the TRC. Nonetheless, we cannot rule out a local heterogeneous water release for plates of low mass, or close from the detachment time, while the gel is under stress. This point certainly deserves more observations with dedicated techniques in a near future.

\section{Conclusion}
We have described the slow aging dynamics of agar plates incubated at constant temperature. The water release leads to the delayed detachment of the gel from the sidewall of the dish. The detachment takes place sooner for plates that display a significant thickness asymmetry, while the detachment time scales quantitatively as a robust function of the gel minimal thickness, which is therefore an excellent candidate to predict the shelflife of commercial plates. To our knowledge, this study is among the first one to use speckle pattern correlation to monitor the gel contraction during syneresis and infer local information on the gel dispacement dynamics. It paves the way for the use of TRC in an industrial context and more generally as a powerful tool to monitor the spontaneous or stress induced syneresis in poroelastic soft media.   
     
\begin{acknowledgments}    
The authors thank S. Manneville \& M. Leocmach for fruitful discussions. This work was partially funded by the BioM\'erieux company.    
\end{acknowledgments}

%\bibliography{rsc}
%

\clearpage

\begin{widetext}

\section*{ Supplemental material to: ``Syneresis and delayed detachment in agar plates"}
\vspace{0.6cm}

\section{Supplemental movies}

{\bf Supplemental movie~1} illustrates the syneresis process and the gel detachment for ten plates from a single batch placed without their lid in a thermoregulated chamber at T=25$^{\circ}$C (gray bullets in Fig.~5 of the main text). Gels progressively release water that evaporates up the moment they detach from the sidewall of the dish, at the exact angular location where their initial thickness was minimal along the plate periphery at the beginning of the experiment. Note that the detachment occurs after a duration ranging from 3.66 to 18.16~hours for this batch and is therefore refered to as {\it delayed}. After the detachment from the wall, the gels shrink significantly in the horizontal direction.

%%%%%%%%%%%%
\begin{figure*}[h]
\centering
\includegraphics[width=0.85\linewidth]{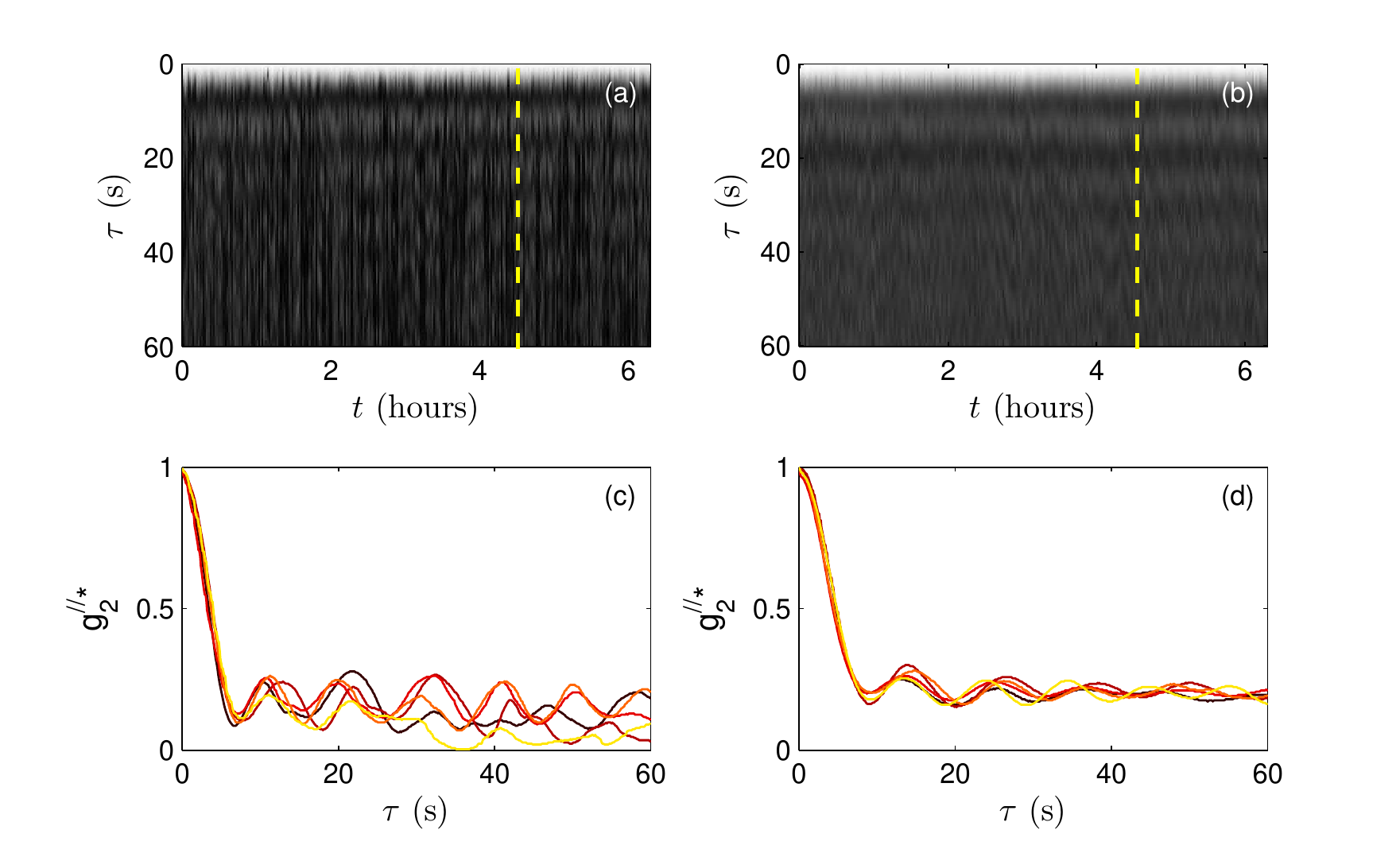}
\caption{ (Color online) (a) and (b) Lag-time temporal diagram of the intensity correlation function $g^{*//}_2(t,\tau)$ coded in grayscale, as a function of the lag time $\tau$ and the experimental time $t$. (c) and (d) Intensity correlation function $g^{*//}_2(t,\tau)$ vs. lag time $\tau$ extracted from (a) and (b) at $t=4.5$~hours for five consecutive one-minute analysis of the speckle correlation. Experiments are performed in parallel configuration (see section 2.3. of the main text) on a gel casted in a plastic Petri dish for (a) and (c), and in a circular dish made of glass of internal radius $39$~mm for (b) and (d). The macroscopic detachment of the gel from the sidewall of the dish occurs at $t=16$~hours for the plastic plate (gel mass $m=26$~g, $e_{min} = 3.74$~mm, $\delta e = e_{max}-e_{min}=0.75$~mm) and  at $t=34$~hours for the glass plate (gel mass $m=22.6$~g, $e_{min}=4.5$~mm and $\delta e=e_{max}-e_{min}=0.7$~mm). 
\label{fig.supfig1}}
\end{figure*} 
%%%%%%%%%%%%

\section{Supplemental figures} 

{\bf Supplemental figure~1} illustrates the lag-time temporal diagram of the intensity correlation function $g^{*\bot}_2(t,\tau)$ coded in grayscale, associated to the dynamics of a commercial gel in a plastic Petri dish [Fig.1~(a)] and of a homemade\footnote[1]{Note that to make sure that time correlation spectroscopy experiments are performed on gels presenting the same mass and thickness in Fig.~1(a) and (b), the gel in the glass dish was prepared in the Lab from a powder provided by BioM\'erieux -- the same powder that is used to prepare the commercial plates otherwise used in this work and in figure~1(a) and (c). The powder is mixed with deionized water and brought to a boil for $15$~min. The hot mixture is then poured into the glass dish and left to cool down at ambiant temperature.} gel in a dish made of glass [Fig.1~(b)]. The glass dish behaves as a non-deformable container which reflects in the temporal evolution of $g^{*\bot}_2(t,\tau)$ which is much smoother than compared to the results obtained in a deformable plastic Petri dish. Indeed, consecutive correlation functions computed over $\tau=1$~min display less variations in the case of the glass dish than in the case of the plastic dish [Fig.1~(c) and (d)]. This experiment demontrates that the stress exerted by the contracting gel on the dish during water evaporation plays a role in the quickly changing dynamics observed in plastic dishes and reported in figure~4(a) and 6 in the main text. Nonetheless, let us emphasize that the fluctuating dynamics of the speckle pattern is still visible in a rigid container and is therefore mainly caused by the gel displacements, which makes it worth studying.
\begin{figure}[h!]
\centering
\includegraphics[width=0.5\linewidth]{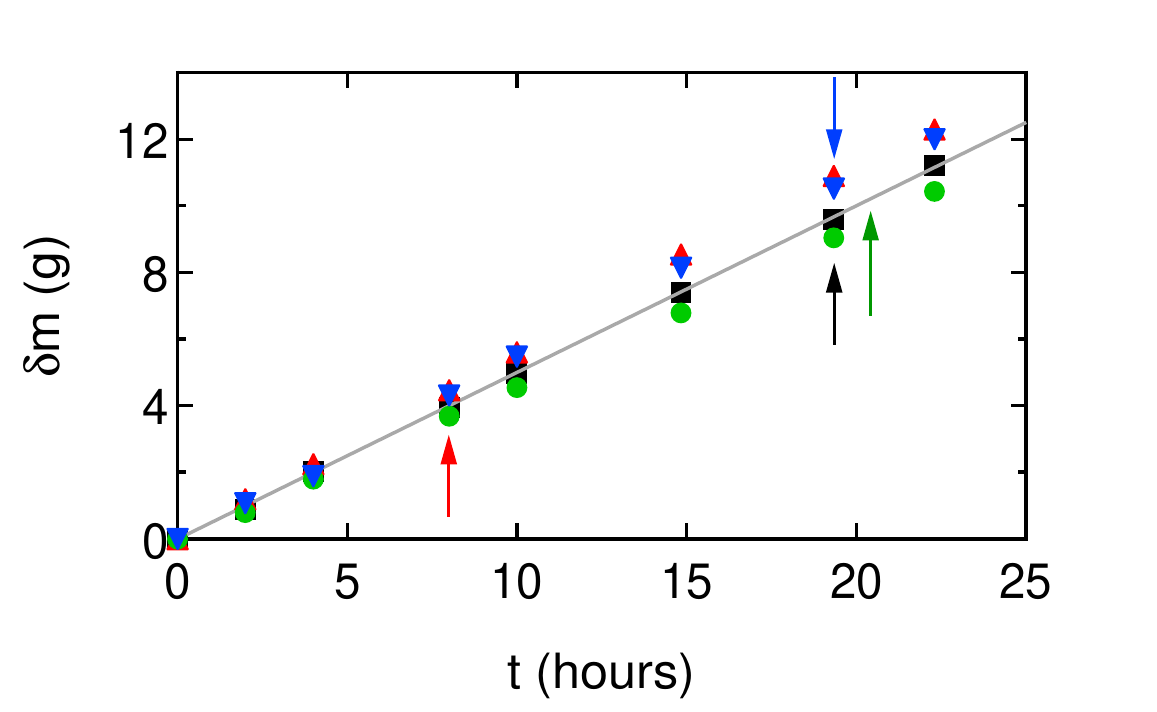}
\caption{(Color online) Mass loss $\delta m \equiv m(t=0)-m(t)$ vs. time $t$ for four different plates taken from three different batches (\textcolor{green}{$\bullet$},\textcolor{red}{$\blacktriangle$},\textcolor{blue}{$\blacktriangledown$},$\blacksquare$). The gray line is the best linear fit over six data sets (two are not shown on the graph for the sake of clarity) with a slope: $\delta \dot m= 0.50 \pm 0.02~$g/hour. Values obtained at other temperatures are given in table~I in the main text. The arrows indicate the detachment time $t^*$ of the gels from the sidewall of the plate.}
\label{fig.dmdt}
\end{figure} 
%%%%%%%%%%%%

{\bf Supplemental figure~2} illustrates the water loss quantified by $\delta m \equiv m(t=0)-m(t)$ for plates maintained in the thermoregulated chamber at $T=20^{\circ}$C. The mass loss increases linearly with time and is not modified by the detachment of the gel. The water-loss rate increases for increasing temperature as reported in table~I in the main text.\\

{\bf Supplemental figure~3} illustrates the evolution of the panoramic 360$^{\circ}$ sideview of an agar plate from the early stage of syneresis up to the detachment. The gel thickness $e$ decreases roughly linearly in time in agreement with the mass loss, while the thickness heterogeneity $\delta e=e_{min}-e_{max}$ remains about constant during the gel thinning (blue and red dashed lines in figure~\ref{fig.profil}). The latter result suggests that the contact area between gel and sidewall, along the whole thickness of the gel (and not particularly along the gel asymetry $\delta e$) is at stake in the building up of stresses that trigger gel micro-displacements. A few hours before gel detachment, red segment lines in figure~\ref{fig.profil} further indicate a more rapid gel thinning in the area of minimum thickness where the growth of a lens shaped meniscus finally leads to sudden gel detachment (last image in figure~\ref{fig.profil}). See also figure~6(e) in the main text.

%\clearpage

%%%%%%%%%%%%
\begin{figure}[!h]
\centering
\includegraphics[angle=-90,width=0.7\columnwidth]{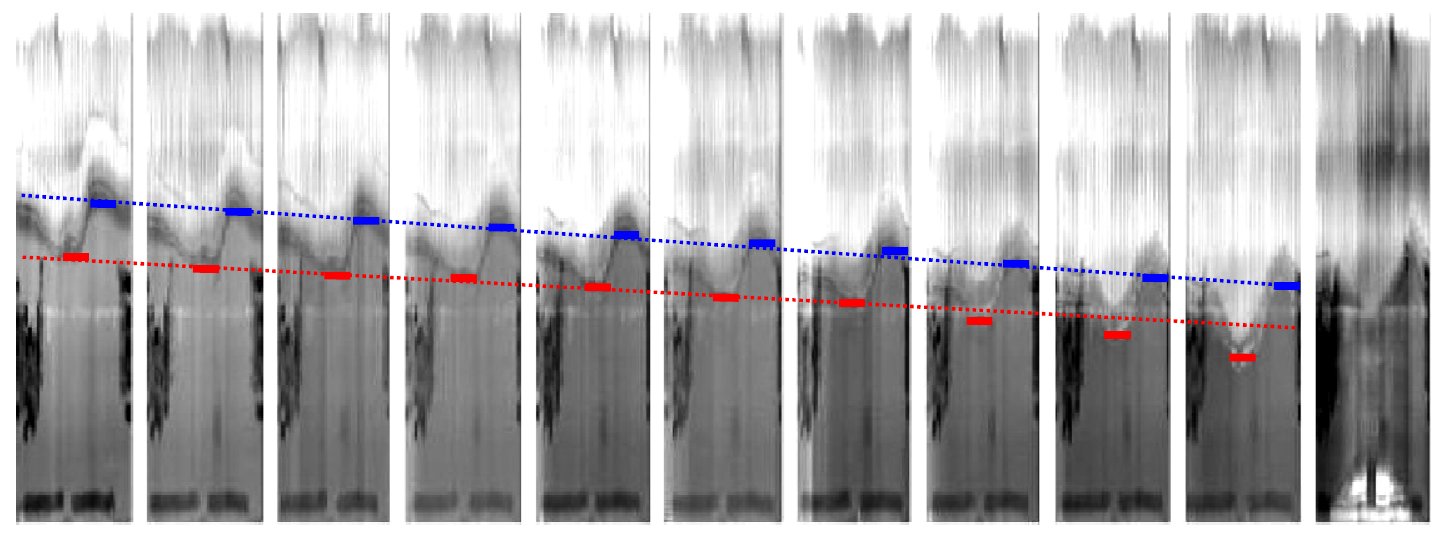}
\caption{Evolution of the 360$^{\circ}$ sideview of a plate during the syneresis process of a plate placed at rest in a thermoregulated chamber (T=25$^{\circ}$C). The first image corresponds to $t=0$, and two successive images are separated by $\Delta t=$1~h. The gel detachment occurs after 9.5~hours, between the last two images. Blue and red segment lines respectively show the time location of the areas of maximal and minimal thickness (blue and red dotted lines are guides for the eye). The initial characteristics of the plate are $e_{min}=3.6$~g, $\delta e=0.7$~mm and $m=22.6$~g.
\label{fig.profil}}
\end{figure} 
%%%%%%%%%%%%

{\bf Supplemental figure~4} displays the lag-time temporal diagram of the intensity correlation function $g^{*\bot}_2(t,\tau)$, obtained in the perpendicular configuration (see section~2.3 in the main text) for a gel that has been carefully detached from the sidewall of the dish before the start of the experiment by means of a cutter blade. This experiment confirms that the second part of the lag-time temporal diagram reported in figure~7(a) in the main text corresponds to the dynamics of a gel that has detached from the sidewall of the dish.
%%%%%%%%%%%%
\begin{figure}[h!]
\centering
\includegraphics[width=0.5\linewidth]{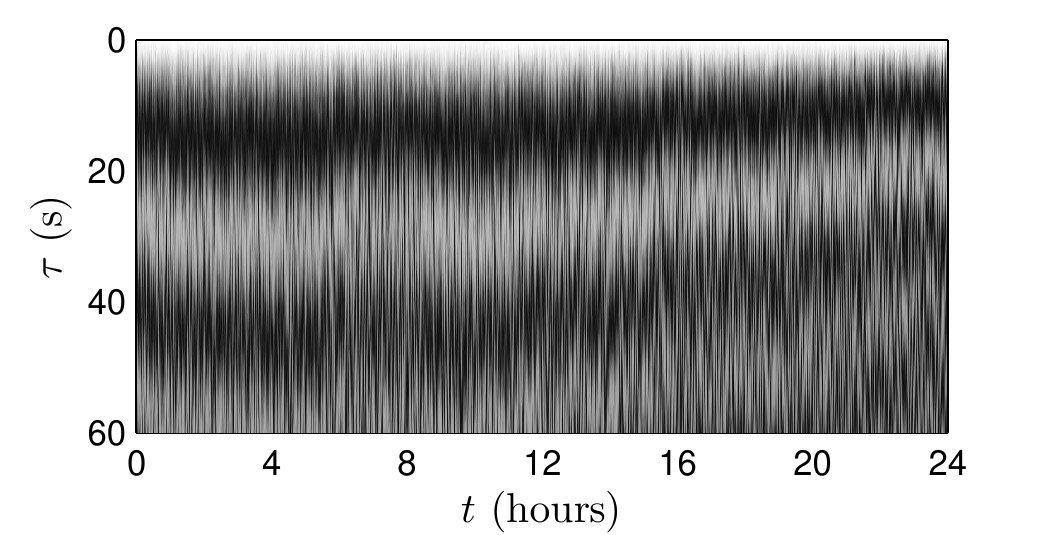}
\caption{Lag-time temporal diagram of the intensity correlation function $g^{*\bot}_2(t,\tau)$, as a function of the lag time $\tau$ and the experimental time $t$. TRC experiments is performed with crossed polarizers and a gel that has been minutely detached from the sidewall of the dish by means of a cutter blade, shortly before the start of the speckle pattern acquisition. The dynamics is very similar to the one reported in figure~7(a) for $t>t^*$ on a standard gel, after the gel has detached.   
\label{fig.cutter}}
\end{figure}

{\bf Supplemental figure~5} shows for three different plates the intensity correlation function $\langle g^{*\bot}_2(t,\tau)\rangle_t$ computed in perpendicular configuration during the early stage of the syneresis, and averaged over the following time interval: 20~min$<t<140$~min - the same as in figure~8 and 9 of the main text. The correlation function decreases as an exponential for $\tau \leq 30$~s, whereas it displays a streched behavior for $\tau\geq 30$~s which can be attributed to the superposition of several phenomena: ($i$) the gel thermal motion, ($ii$) the micro-displacements of the gel inside the dish while the gel thins but remains macroscopically in contact with the sidewall, and ($iii$) local changes in the gel surface topography or in the orientation of the air/gel interface induced by the gel micro-displacements. 
%%%%%%%%%%%%
\begin{figure}[h]
\centering
\includegraphics[width=0.5\columnwidth]{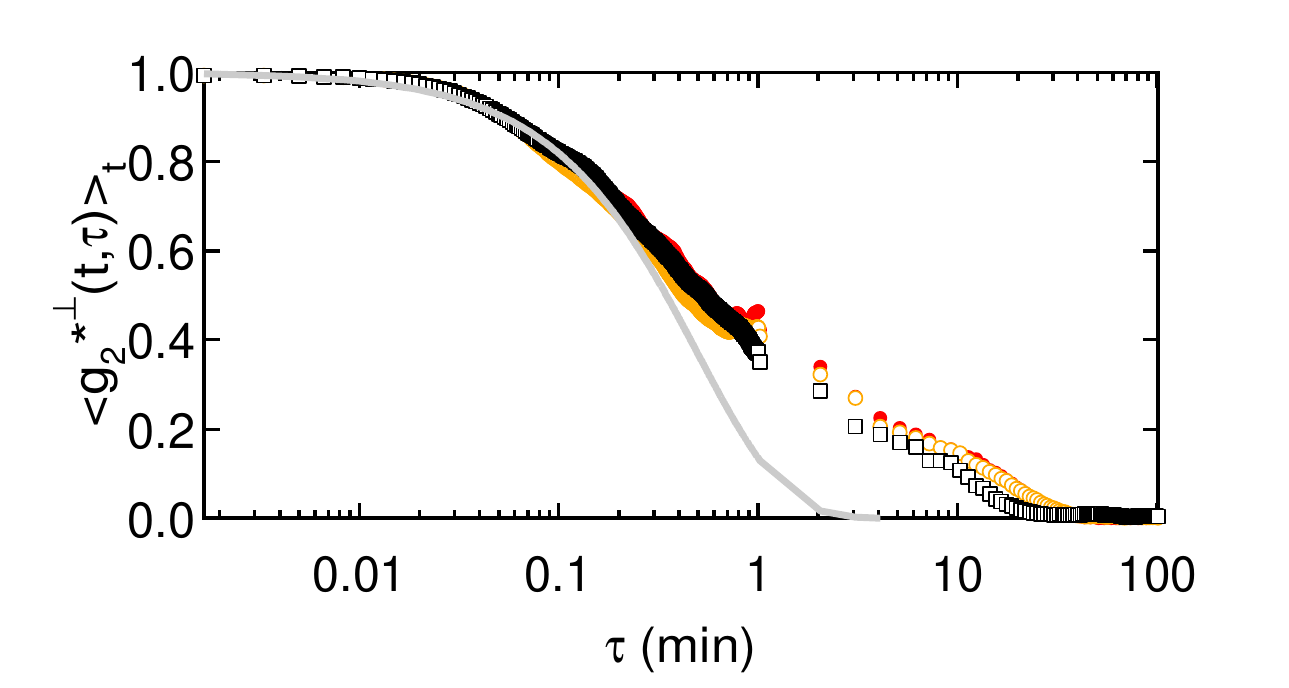}
\caption{(Color online) Averaged correlation function $\langle g^{*\bot}_2(t,\tau)\rangle_t$ vs. $\tau$ for three different plates (same symbols as in fig.~8 in the main text). The gray line corresponds to a decreasing exponential function over a characteristic timescale $\tau_0$=30~s. Note that here, $g^*_2(t,\tau)$ is computed at the maximum frame rate of 10~Hz over the lag time $\tau = 1$~min, while for $\tau >1$~min, $g^*_2(t,\tau)$ is only computed every minute. 
\label{fig.corFunction}}
\end{figure} 
%%%%%%%%%%%%

\end{widetext}

 \end{document}